\documentclass[twocolumn, floatfix, prb, superscriptaddress, aps,showpacs, longbibliography]{revtex4-2}
\usepackage{graphicx,amsmath,amssymb, color, bm,mathtools}
\usepackage{mathrsfs}
\usepackage{nicefrac}
\usepackage{fancyhdr, lipsum}
\usepackage[titletoc,title]{appendix}
\usepackage[dvipsnames]{xcolor}
\usepackage[normalem]{ulem}

\newcommand{\be}{\begin{equation}}
\newcommand{\ee}{\end{equation}}

\newcommand{\ba}{\begin{eqnarray}}
\newcommand{\ea}{\end{eqnarray}}

\newcommand{\abs}[1]{\left|#1\right|}
\newcommand{\avg}[1]{\langle#1\rangle}

\newcommand{\LLL}{\mathcal{P}_\mathrm{LLL}}
\newcommand{\lB}{\ell_{B}}

\usepackage{orcidlink}
\newcommand{\oovv}[1]{\overline{\overline{#1}}}
\newcommand{\ov}[1]{\overline{#1}}

\begin{document}

\title{Exploring the nature of the emergent gauge field in composite-fermion metals: A large-scale microscopic study}

\author{Amogh Anakru\orcidlink{0000-0002-9818-9483}}
\thanks{These authors contributed equally to this work.}
\affiliation{Department of Physics, 104 Davey Lab, Pennsylvania State University, University Park, Pennsylvania 16802, USA}

\author{Mytraya Gattu\orcidlink{0000-0001-6994-389X}}
\thanks{These authors contributed equally to this work.}
\affiliation{Department of Physics, 104 Davey Lab, Pennsylvania State University, University Park, Pennsylvania 16802, USA}
\email{mvg6042@psu.edu}

\author{Ajit C. Balram\orcidlink{0000-0002-8087-6015}}
\affiliation{Institute of Mathematical Sciences, CIT Campus, Chennai, 600113, India}
\affiliation{Homi Bhabha National Institute, Training School Complex, Anushaktinagar, Mumbai 400094, India}

\author{Xiao-Chuan Wu\orcidlink{0000-0002-5503-3630}}
\affiliation{Kadanoff Center for Theoretical Physics, University of Chicago, Chicago, IL 60637, USA}
\affiliation{Department of Physics, Princeton University, Princeton, NJ 08544, USA}

\author{Prashant Kumar\orcidlink{0000-0002-5800-2768}}
\affiliation{Department of Physics, Indian Institute of Technology Bombay, Mumbai, MH 400076, India}

\author{Zhen Bi\orcidlink{0000-0003-0351-3963}}
\affiliation{Department of Physics, 104 Davey Lab, Pennsylvania State University, University Park, Pennsylvania 16802, USA}

\author{J. K. Jain\orcidlink{000-0003-0082-5881}}
\affiliation{Department of Physics, 104 Davey Lab, Pennsylvania State University, University Park, Pennsylvania 16802, USA}
\email{jkj2@psu.edu}

\begin{abstract}
Field theories of the composite-fermion (CF) metal model it as a Fermi sea of composite fermions coupled to an emergent gauge field. Within a random phase approximation, these theories predict that the Landau damping of the gauge field resulting from its coupling to the low-energy, long-wavelength CF particle-hole excitations modifies the electrons' density-density correlation function related to the static structure factor $S(q)$ at wave vector $q$. This produces a non-analytic correction $\propto q^{3}\ln q$ to $S(q)$ (with the magnetic length $\lB=1$). Thanks to the recently developed quaternion formulation for Jain-Kamilla projection of CF wave functions, the evaluation of $S(q)$ from the accurate microscopic theory of composite fermions has now become possible for systems containing as many as $N=900$ CFs, which enables a reliable determination of the small-$q$ behavior of $S(q)$. We study CF metals corresponding to electrons at Landau level filling factors $\nu=1/2$ and $1/4$, and for completeness, also of bosons at $\nu=1$ and $1/3$. In the $q\rightarrow0$ limit, our microscopic calculation reveals a $q^{3}$ term in $S(q)$ of the CF metals rather than $q^{3} \ln q$. This behavior is well-predicted by a model of a non-interacting Fermi sea of dipolar CFs, which also obtains its coefficient accurately.
\end{abstract}

\maketitle

\textit{Introduction.} Electrons in metals typically form Fermi liquids (FLs). One obtains a simple Fermi sea when interactions are switched off, with the occupation number dropping discontinuously from 1 to 0 across the Fermi wave number $k_F$. The central tenet of the FL theory is that when the interparticle interaction is turned on, the jump at $k_F$ becomes smaller but remains finite, resulting in long-lived electron-like quasiparticles sufficiently close to the Fermi energy~\cite{Mahan00, Giuliani08}. A strong enough interaction, however, can cause this jump to vanish, thereby yielding what is termed a non-Fermi liquid (NFL) \cite{Halperin93, Nayak94nfl,lee_nagaosa_wen2006, senthil2008,sachdevQPT}.

Of interest to us here is the system of interacting electrons in the half-filled lowest Landau level (LLL). No Fermi sea exists in the absence of interactions—the system is macroscopically degenerate. Turning on a repulsive interaction, no matter how small, has the non-perturbative effect of generating composite fermions (CFs)~\cite{Jain89}, namely bound states of electrons and an even number ($p$) of vortices. At even-denominator fillings $\nu=1/p$, the CFs experience zero effective magnetic field and form a Fermi-sea-like state. Here, the interaction creates a Fermi-sea-like state where there was none to begin with~\cite{Jain89, Halperin93}. This CF metal is one of the most dramatic examples of an NFL of electrons, whose low-energy excitations are topologically distinct from electrons. However, one may ask: To what extent does this state resemble a regular FL of CFs?

Experimentally, the CF metal at $\nu=1/2$ has been investigated extensively~\cite{Shayegan20} and displays many familiar hallmarks of an ordinary FL: it has a sharply defined Fermi wave vector which has been measured with great precision~\cite{Kamburov14b}; it obeys the Luttinger area rule~\cite{Luttinger60a, Luttinger60b} to high accuracy and with little sensitivity to interaction strength (or LL mixing)~\cite{Balram15b, Geraedts16, Balram17, Hossain20}; it exhibits quantum oscillations near half filling~\cite{Du94}; and it hosts high-order Jain states corresponding to the integer quantum Hall effect (IQHE) of CFs~\cite{Chung21}.

In the field-theoretic descriptions of Halperin-Lee-Read (HLR)~\cite{Halperin93, Simon98, Halperin03}, the CF metal is distinguished from regular FLs in that, in addition to the inter-CF interaction, the CFs are also coupled to an emergent Chern-Simons gauge field arising from flux attachment~\cite{Lopez91}. This field interacts strongly with the continuum of low-energy, long-wavelength CF particle-hole excitations available in the compressible CF metal. Within the random-phase approximation (RPA), such coupling causes a Landau damping of the gauge field and modifies the long-wavelength density-density response. The Landau-damped gauge field dresses the CFs, destroying the long-lived quasiparticles at the Fermi surface, eliminating the discontinuous jump in the CF occupation number at $k_{F}$~\cite{Nayak94nfl, sachdevQPT}. Another consequence of this Landau-damped gauge field is that the ground-state static structure factor $S(\bm{q})$ contains a non-analytic $q^3\ln q$ term (the magnetic length $\lB = \sqrt{\hbar c/eB}$ at magnetic field $B$ is set to unity hereafter):
\begin{equation}\label{eq:hlr-predict}
    S(\bm{q}) = \frac{q^{2}}{2} + \frac{2-\eta}{2\pi}k_{F}q^{3}\ln\!\left(\frac{q_{0}}{q}\right) + \mathcal{O}(q^4),
\end{equation}
where $U(q) \sim q^{-\eta}$, $0\!\leq\!\eta\!\leq\!2$, is the interaction between fermions. This is in contrast to $S(\bm{q})$ of 
incompressible fractional quantum Hall effect (FQHE) states, for which $S(\bm{q})$ is an analytic function of  $q^2=|\bm q|^2$.
The objective of this work is to determine whether the $S(\bm{q})$ obtained from an accurate microscopic theory displays this behavior. Throughout this work, we consider only the isotropic CF metal, for which $S(\bm{q})$ reduces to a function of $q$ alone, and can thus be denoted $S(q)$. 

Earlier studies have investigated this question. Balram and Jain~\cite{Balram17} analyzed the CF metal using CF wave functions on the sphere, but, limited to $N\!\leq\!81$ particles, this study could not access the thermodynamic limit of $S(q)$ for $q \lesssim 1$, leaving the small-$q$ behavior unresolved. Kumar and Haldane~\cite{Kumar22}, using density matrix renormalization group (DMRG) in cylindrical geometry, did not find evidence for a $q^3\ln q$ term; however, they argued that this was an artifact of the quasi-one-dimensional setting. In the present work, we carry out a large-system, fully two-dimensional microscopic study of $S(q)$ for the CF metal, enabled by the recently developed quaternion formulation of the Jain--Kamilla projection~\cite{Gattu25} for the CF wavefunctions in the spherical geometry. This method allows us to access systems of up to $N=900$ particles (and possibly more) and obtain converged thermodynamic results for $S(q)$ down to $q\gtrsim 0.10$.

\textbf{Microscopic calculation of $S(q)$.} To extract the bulk physics of the CF metal, we work in Haldane's spherical geometry~\cite{Haldane83}.
In this setup, particles are confined to the surface of a sphere of radius $R$ enclosing a magnetic monopole of strength $2Q$ at its center. The monopole generates a uniform radial magnetic field $B$ over the surface, with a total magnetic flux of $2Q\phi_0$, where $\phi_0=hc/e$ is the magnetic flux quantum. The sphere's radius is related to the monopole strength by $R=\sqrt{Q}$.

The single-particle eigenstates are the monopole harmonics $Y_{Q,l,m}(\theta,\phi)$~\cite{Wu76, Wu77}, which are organized into shells of angular momentum $l$. These shells correspond to LLs and are labeled by the principal quantum number $n_l=0,1,2,\dots$, such that $l=Q+n_l$. Each shell contains $2l+1$ degenerate orbitals distinguished by the azimuthal quantum number $m \in \{-l, -l+1, \dots, l\}$. The monopole harmonics are conveniently expressed as polynomials of the spinor variables $u=\cos(\theta/2)e^{i\phi/2}$ and $v=\sin(\theta/2)e^{-i\phi/2}$. In the LLL, which has $n_l=0$ ($l=Q)$, the eigenfunctions take the simple form $Y_{Q, Q,m} \propto u^{Q+m}v^{Q-m}$. 

We next turn to the construction of the many-body CF wave functions. Consider $N$ particles confined to the LLL in the presence of a monopole of strength $2Q$. In the CF construction, each particle binds $p$ vortices, producing CFs that experience a reduced monopole strength $2Q^{\ast}=2Q-p(N-1)$. In this effective field, the CFs occupy their own LL-like shells, called the $\Lambda$ levels ($\Lambda$Ls). The wave function of $n$ filled $\Lambda$Ls describing the Jain state at $\nu=n/(pn+1)$ is constructed as~\cite{Jain89, Jain07}
\begin{equation}
\Psi({\theta_i,\phi_i};Q) = \LLL\Phi_{n}({\theta_i,\phi_i};Q^{\ast})\Phi_1^p,
\label{eq:cf-wavefunction}
\end{equation}
where $\Phi_{n}$ is the Slater determinant for $N$ electrons filling the lowest $n$ LLs at monopole strength $2Q^{\ast}=(N/n-n)/2$, the Jastrow factor $\Phi_1^p=\prod_{i<j}(u_i v_j - u_j v_i)^p$ attaches $p$ vortices to each electron, $\LLL$ represents projection of the wave function into the LLL. Thus, we have $Q=(N/n-n)/2+p(N-1)/2$, and $\nu=\lim_{N\rightarrow\infty}N/2Q=n/(pn+1)$. 

For electrons, $p$ must be an even integer, ensuring that the wave function is antisymmetric under the exchange of two particles. For completeness, we will also consider odd-integer $p$, in which case the wave function describes charged bosons in the LLL, and is interpreted as $n$ filled $\Lambda$Ls of CFs formed from the binding of a boson and $p$ vortices. 

We construct the wave function for the CF metal at $\nu=1/p$ by filling the lowest $n$ $\Lambda$ levels at $Q^{\ast}=0$ (corresponding to $Q=p(N-1)/2$) with $N=n^{2}$ CFs~\cite{Rezayi94}, and taking the limit $n\rightarrow \infty$. These states are gapped for finite $n$, but the gap vanishes as $n\to\infty$~\cite{Gattu25}, yielding the compressible Fermi sea state. We note that for charged bosons at $\nu=1$, the true ground state for the contact or the Coulomb interaction is likely a FQHE state in the bosonic Moore-Read phase~\cite{Moore91} rather than the CF metal~\cite{Wilkin00, Cooper01, Regnault03, Chang05b, Viefers08, Cooper20, Sharma24}; however, the CF metal phase considered here can presumably be stabilized by some other interaction~\cite{Wu15}. For $p=2,3,4$ corresponding to $\nu=1/2,1/3,1/4$, the CF metal states are known to be excellent representations of the Coulomb ground states~\cite{Dev92a, Wu93, Rezayi94, Jain07, Liu20}. 

A crucial step is the implementation of the LLL projection. The conventional Jain--Kamilla (JK) method~\cite{Jain97, Jain97b}  scales algebraically with $N$ as $\mathcal{O}(N^{3})$ and allows, in practice, evaluation of wave functions with up to $\sim 9$ filled $\Lambda$Ls. As noted above, the finite size corrections here are significant for $q\lB \lesssim 1.0$. Critical to the present work is a recently introduced method that exploits a quaternion representation of the wave functions~\cite{Gattu25}, which enables the study of systems with up to $\gtrsim 30$ filled $\Lambda$Ls and obtains thermodynamic limits for $q\lB \gtrsim 0.1$.

We now turn to the calculation of the static structure factor $S(q)$ using these wavefunctions. The static structure factor of a state $\lvert 0 \rangle$ with $N$ particles is defined as 
\begin{equation} \label{eq:ssf-plane-def}
S(\mathbf{q}) = \frac{1}{N}\langle 0\lvert \rho^{\dagger}_{\mathbf{q}}\rho_{\mathbf{q}} \rvert 0 \rangle - N\delta_{\mathbf{q},0};\;\;
\rho_{\mathbf{q}} \equiv \sum_{j=1}^{N} e^{i \mathbf{q}\cdot \mathbf{r}_j}.
\end{equation}
On the sphere, the plane-wave basis $e^{i\mathbf{q}\cdot \mathbf{r}}$ is replaced by spherical harmonics $Y_{l,m}(\theta,\phi)$, giving~\cite{He94} 
\begin{equation}
S_{l,m} = \frac{4\pi}{N}\langle 0 \lvert \rho^{\dagger}_{l,m}\rho_{l,m} \rvert 0 \rangle - N\delta_{l,0};\;
\rho_{l,m} = \sum_{j=1}^{N} Y_{l,m}(\theta_j,\phi_j).
\end{equation}
For uniform states $S_{l,m}$ is independent of $m$, so we denote it as $S_l$. We evaluate $S_l$ using the Metropolis--Hastings--Gibbs Monte Carlo algorithm~\cite{Kamilla97, Binder10} with the wave functions as the sampling weight, with $4\times 10^6$ samples across $50$-$100$ independent runs. 

\begin{figure}    \includegraphics[width=\columnwidth]{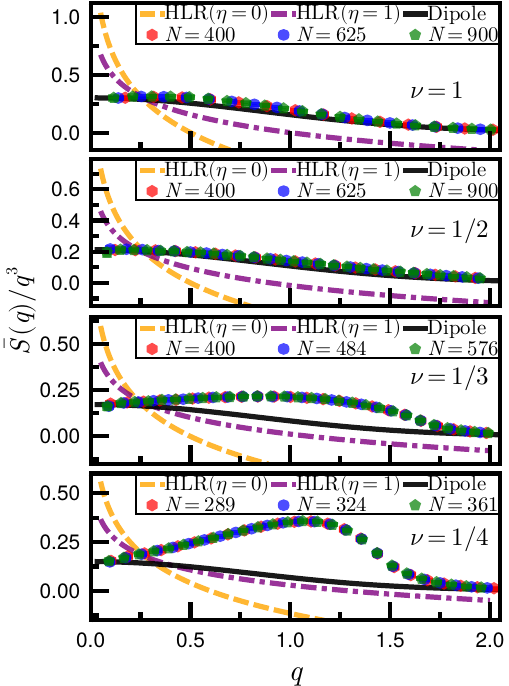}
    \caption{The projected static structure factor $\overline{S}(q)$ is evaluated for the CF metal states at $\nu=1$ and $\nu=1/3$ for bosons and at $\nu=1/2$ and $\nu=1/4$ for fermions in the lowest Landau level, using composite-fermion (CF) trial wave functions at zero effective magnetic field. To highlight the small-$q$ behavior, we plot $\overline{S}(q)/q^{3}$ versus $q$. The points denote CF results obtained for several large system sizes ($N$), with vertical error bars indicating Monte Carlo uncertainties (which are smaller than the symbol sizes). The solid light blue and purple lines show the Halperin-Lee-Read (HLR) prediction, $\overline{S}(q)=(2-\eta)/(2\pi)k_{F} q^{3}\ln(q_{0}/q)$, for $\eta=0$ (short-range interaction) and $\eta=1$ (Coulomb interaction) respectively at the level of the random phase approximation (RPA). Here, $\eta$ parametrizes the long-wavelength form of the electron interaction, $U(q)\sim q^{-\eta}$, and $q_{0}$ is obtained from a best fit for $q \leq 0.50$. The solid yellow line shows $\bar{S}(q)$ obtained in a model of noninteracting dipolar CFs (see text), which behaves as $\overline{S}(q)\sim(2k_{F}/3\pi)q^{3}$ in the limit of $q\rightarrow 0$. The wave number $q$ is measured in units of $1/\ell_{B}$, where $\ell_{B}=\sqrt{\hbar c/eB}$ is the magnetic length.}
    \label{fig:projected-ssf-projected-CF metal}
\end{figure}

Girvin-MacDonald-Platzman (GMP)~\cite{Girvin85, Girvin86} introduced a projected static structure factor $\overline{S}(\bm{q})$ by replacing in Eq.~\eqref{eq:ssf-plane-def}, $\rho_{\bm{q}} \rvert 0 \rangle$ by $\LLL \rho_{\bm{q}} \rvert 0 \rangle$. It is related to $S(q)$ as 
\begin{equation}
\label{eq:projected-sff-plane-def}
  \overline{S}(\bm{q})=S(\bm{q})-[1-e^{-q^2/2}],  
\end{equation}
where the last term in square brackets is the inter-LL contribution, which in the $q\rightarrow 0$ limit gives the universal Kohn contribution of $q^{2}/2$~\cite{Kohn61} for any uniform state~\cite{Girvin86, Dora24, Wu24}. We plot below $\overline{S}(q)$ to isolate the behavior specific to the CF metal.

Two technical points have proved important for obtaining the thermodynamic behavior. First, we relate the planar wave vector $q$ to the spherical angular momentum $l$ as $q=\sqrt{l(l+1)}/R$ rather than the commonly used $q=l/R$~\cite{Haldane85a, Simon94a, He94, Jain07}. While both are adequate at large $q$, only with the former definition of $q$ does the $S(q)$ calculated in the spherical geometry converge to the planar results for the IQHE states in the limit $q \rightarrow 0$. We find that also for the FQHE states, $q=\sqrt{l(l+1)}/R$ provides a much better convergence as a function of $N$. Appendix~\ref{app: S_L_plane_sphere_IQHE} contains further discussion of this issue. 
Second, $S(q)$ at even-$l$ exhibits much faster convergence to the thermodynamic behavior as a function of $N$ than that at odd-$l$ values, as shown in Appendix~\ref{app:even-odd}. We therefore show $S(q)$ only for $q$ corresponding to even-$l$ values. Note that $l=1$ is absent anyway since $\overline{S}_{l=1}$ vanishes identically for any uniform state in the LLL~\cite{Dev92}. 

{\bf Microscopic vs. field theories.} 
Fig.~\ref{fig:projected-ssf-projected-CF metal} presents the $\overline{S}(q)$ obtained from our microscopic calculations at $\nu=1,\;1/2,\;1/3,$ and $1/4$. For each filling, the calculated $\overline{S}(q)$ values for different $N$ fall on a single curve, thus demonstrating convergence to the thermodynamic limit. With the largest systems studied, we can resolve $\overline{S}(q)$ down to $q\ell_{B}\approx 0.1$.

Alongside the microscopic results, we have plotted the HLR prediction [see Eq.~\eqref{eq:hlr-predict}] for $\eta=0$ and $\eta=1$; these correspond respectively to electrons interacting via a generic short-range interaction~\cite{Wu24} and the Coulomb interaction, with $q_{0}$ chosen to minimize the deviation from our microscopic $\overline{S}(q)$ at small-$q$. Our microscopic results clearly show that for the CF metals at $\nu=1,1/2,1/3,$ and $1/4$, in the small-$q$ limit $\overline{S}(q)$ behaves as $q^{3}$ and not as  $q^{3}\ln q$ as predicted by the field theory.
  
The HLR theory has been enormously successful in predicting and describing many experiments; an overview can be found in the review by Halperin~\cite{Halperin20}. Yet, at the level of RPA, requiring a physically sensible response fixes $\overline{S}(q)$ to behave as $q^{3}\ln q$ as $q\to 0$. The Son-Dirac effective field theory also makes similar predictions. (See Appendix~\ref{app: SSF_field_theories} for details.)

We mention some possible sources of this discrepancy. In the HLR theory, at the mean-field level, the CFs are defined by flux attachment, implemented through a singular gauge transformation. The mean field state $\prod_{j<k}[(z_j-z_k)/|z_j-z_k|]^p\Phi_n(B)$ captures the phase factors but does not build good repulsive correlations between electrons (its modulus is the same as that of the IQHE state $\Phi_n$, and thus the probability of two particles separated by a distance $r$ approaching one another vanishes as $r^2$) and also has significant occupation of higher LLs~\cite{Kamilla97, Balram16b}. Because the mean field state is far from the actual state, we expect that it will be difficult for perturbation theory to account for all the fluctuations about the mean field that restore the true ground state. 
The RPA treatment of HLR has been justified 
in a weak-coupling perturbative renormalization-group approach~\cite{Nayak94nfl, mross2010} for the long-ranged Coulomb interaction. However, when the interactions are short-ranged, more sophisticated approaches are required. Often the RPA can be justified in a large-$N$ approximation, but here the large-$N$ approximation itself is uncontrolled~\cite{sung-sik-lee-2009}, and more elaborate generalizations of the large-$N$ expansion are needed~\cite{mross2010, damia2019_matrixLargeN, sachdev_criticalFS_1, sachdev_criticalFS_2}. These controlled limits may not capture the physics of the CF metals. Furthermore, we note that the HLR theory does not explicitly implement LLL projection, and therefore it is unclear to what extent it would capture the physics of the LLL projected CF metals.

We are also assuming that the CF wave function of Eq.~\eqref{eq:cf-wavefunction} provides a faithful approximation for the actual Fermi sea. The wave functions for the $\nu=n/(pn\pm 1)$ for $p=2$ and $n=1$, $2$ and $3$ are almost the same as the exact Coulomb ground states obtained from exact diagonalization~\cite{Jain97, Jain97b, Jain07, Balram13, Balram21b}, and the CF Fermi sea (CFFS) wave function has also been found to be highly accurate in the torus geometry~\cite{Rezayi00}. This suggests that the CF wave function is accurate also for the CF metal, but we cannot rule out the possibility that the wave function may not capture the long-distance behavior of the actual Coulomb CF metal. 

\textbf{Interpretation in terms of free CFs.} In contrast to the HLR theory, in the Jain state  $\LLL\prod_{j<k}(z_j-z_k)^p\Phi_n(B^{*})$ a CF is defined as an electron bound to {\it vortices}, and repulsive correlations are evident before LLL projection (the probability of two particles approaching one another vanishes as $r^{2p+2}$). Ref.~\cite{Balram13} finds that the widths of the low lying $\Lambda$L bands (relative to the separation between them) decrease as we go to higher $n$, suggesting that the inter-CF interaction actually becomes weaker as we approach $\nu=1/2$. This justifies why the CFs defined in the microscopic theory may be taken as weakly interacting. 

In view of these remarks, it is natural to ask whether the long-wavelength density-density response of the CF metal can be captured in a model of noninteracting CFs. We now show that such a model exists, and that it not only reproduces the $q^{3}$ behavior of $\bar{S}(q)$ but also yields the correct coefficient of the $q^{3}$ term.

Let us consider the CF Fermi sea and ask what happens when we apply the projected electron density operator $\ov{\rho}_{\bm{q}}$ to it. The action of the electron annihilation operator $\Psi^{\rm electron}(\eta)$ on $\Psi(\{ z_1,\cdots,z_N\})$ produces $\Psi(\{ z_1,\cdots,z_{N-1},\eta\})$, which has a very complicated interpretation in terms of CFs and is a high energy state with many CF holes in various $\Lambda$Ls. The action of the electron creation operator, followed by the LLL projection, similarly produces a complicated state~\cite{Peterson05, Gattu24, Pu23a}. Thus, one expects a single electron-hole excitation created by $\ov{\rho}_{\bm{q}}$ to be a complicated state consisting of many CF particle-hole excitations. Remarkably, and surprisingly, it was shown in Ref.~\cite{Balram24} that a single electron-hole excitation can be expressed as a linear superposition of \textit{single CF particle-hole excitations}. In other words, the state $\ov{\rho}_{l}\LLL \Phi_1^p \Phi_n$ can be written as a linear superposition of states of the form $\LLL \Phi_1^p \rho_{l,n_1,n_2} \Phi_n$ where $\rho_{l,n_1,n_2}$ creates a particle hole pair at angular momentum $l$ by exciting an electron from an occupied $\Lambda$L $n_1$ to an unoccupied $\Lambda$L $n_2$. 

Let us schematically define \textit{CF} creation and annihilation operators $c^\dagger$ and $c$, such that $c^\dagger_{k_1}c_{k_2}\LLL \Phi_1^p  \Phi_n \sim \LLL \Phi_1^p \Phi^{k_{1},k_{2}}_n$, where $\Phi_{n}^{k_{1}, k_{2}}$ is an electron-hole exciton state at $\nu=n$ IQH state with a hole in $k_{2}$ and an electron in $k_{1}$. Encouraged by the result of~\cite{Balram24}, we postulate a representation of the electron density operator $\bar{\rho}(q)$ as a superposition of single CF excitons, i.e. as a bilinear in the $c^\dagger_{k_1}$, $c_{k_2}$ operators. We also suppose that the CFs form a non-interacting Fermi sea and that the corresponding $\bar{S}(q)$ vanishes faster than $q^2$ at small $q$, as implied by Kohn's theorem~\cite{Murthy03}. Finally, we require that the operators $e^{q^2/4}\bar{\rho}(q)$ obey the GMP algebra. In Appendix~\ref{app:GMP_MOTIVATION}, we show that the above requirements uniquely fix our postulated density operator to be
\begin{equation}\label{eq:rho-small-q}
{\overline{\rho}}(\bm{q})=e^{-q^2/4}\sum_{\bm{k}} i\ell_{\rm B}^{2}(\bm{q}\times \bm{k})c^\dagger_{\bm{k}+\bm{q}} c_{\bm{k}},
\end{equation}
which produces (see Appendix~\ref{app: dipole_Fermi_sea}) \begin{equation}\overline{S}(q) = \frac{2k_F}{3\pi}q^3 + \dots, \;\;\; k_{\rm F}=\sqrt{2\nu}
\end{equation}
in the small-$q$ limit. As seen in Fig.~\ref{fig:projected-ssf-projected-CF metal}, this is in excellent agreement with the $\overline{S}(q)$ obtained in our microscopic calculation. Significantly, even the coefficient of this term is very close to the value obtained from the microscopic calculation. 

Eq.~\eqref{eq:rho-small-q} was earlier shown to arise from a `dipole interpretation' where a dipole moment $\lB^2\hat{z}\times\bm{k}$ is assigned to each CF~\cite{Read98}. In real space, the charge density in Eq.~\eqref{eq:rho-small-q} is just the divergence of such a dipole density $\bm{\rho}_d(\bm{q}) = \sum_{\bm{k}}\lB^2\hat{z}\times\bm{k}c^\dagger_{\bm{k}+\bm{q}} c_{\bm{k}}$. Eq.~\eqref{eq:rho-small-q} is also the $q\to 0$ limit of
\begin{equation}
\label{eq:GMP_charge_density_operator}
\overline{\rho}(\bm{q}) = e^{-q^2/4}\sum_k 2i\sin((\bm{q}\times\bm{k})/2)c^\dagger_{\bm{k}+\bm{q}}c_{\bm{k}}
\end{equation}
derived by Murthy and Shankar in a Hamiltonian approach~\cite{Murthy07} and by Dong and Senthil in a non-commutative field theory of CFs~\cite{Dong20}, building on a parton formulation~\cite{Pasquier98, Read98, Ma22} of the CF metal. These approaches apply only to the bosonic metal at $\nu=1$~\cite{Read98}, though some proposals extend it to $\nu=1/2$ \cite{Gocanin2021Dirac, Predin2023DipoleHalf, GreenThesis02}. We evaluate $\overline{S}(q)$ with this form and a non-interacting, circular Fermi sea with $k_F$ set by the filling $\nu=1/p$. The resulting $S(q)$ is compared to the microscopic calculation for $p=1,2,3,4$ in Fig.~\ref{fig:projected-ssf-projected-CF metal}. (We note that using Eq.~\eqref{eq:rho-small-q} produces essentially identical $S(q)$.) This produces excellent agreement with the microscopic calculation at small and large $q$; for $\nu=1/2$, the agreement is quite good over the entire range of $q\ell_{\rm B}$. (For $\nu=1/3$ and $\nu=1/4$, the $\overline{S}(q)/q^{3}$ versus $q$ plots exhibit a positive linear component, which can be attributed to the parton graviton mode supported by these states~\cite{Balram21d, Nguyen22, Bose25}.) Kumar and Haldane \cite{Kumar22} have also used an expression similar to Eq.~\eqref{eq:rho-small-q} at $\nu=1/2$. In that work, while the dynamical structure factor of electrons was found to be consistent with this prescription, emergence of the $q^3\ln q$ term of Eq.~\eqref{eq:hlr-predict} at large sizes could not be ruled out. By contrast, our present study - effectively in the thermodynamic limit - does not find the contribution of the Landau-damped emergent gauge field to the static structure factor.

In summary, we have presented a microscopic computation of the static structure factor for various CF metals and found that the results are in disagreement with the predictions of the effective field theories of CFs in the small $q$ limit. We then show that the behavior is accurately described by modeling the CF metal as a non-interacting CFFS. In the future, it would be interesting to investigate other quantities from a microscopic approach---such as the entanglement entropy~\cite{Shao15, Voinea25}, and fluctuations of conserved quantities generalizing the results of~\cite{Wu24_CFS, Wu24}---to better understand the role of the emergent gauge field.

\textit{Note added}: During the preparation of this manuscript, Ref.~\cite{chen2025probing} appeared which also considers similar issues. 

\begin{acknowledgments}
We thank T. Senthil for insightful discussions. M.G. and J.K.J. acknowledge support in part by the National Science Foundation under Grant No. DMR-2404619. A.A. and Z.B. acknowledge the support from NSF under award number DMR-2339319. A.C.B. acknowledges the Science and Engineering Research Board (SERB) of the Department of Science and Technology (DST) for financial support through the Mathematical Research Impact Centric Support (MATRICS) Grant No. MTR/2023/000002. X.W. was supported in part by the Simons Collaboration on Ultra-Quantum Matter, which is a grant from the Simons Foundation (651442), and the Simons Investigator Grant (566116) awarded to S. Ryu. P.K. was supported in part by Prime Minister Early Career Research Grant from Anusandhan National Research Foundation project no. ANRF/ECRG/2024/002274/PMS, and in part by Seed Grant from IIT Bombay. The authors of this work recognize the Penn State Institute for Computational and Data Sciences (RRID:~SCR\_025154) for providing access to computational research infrastructure within the Roar Core Facility (RRID:~SCR\_026424).
 \end{acknowledgments}
\bibliography{biblio_fqhe.bib}
\newpage

\appendix
\section*{Supplementary Material}

In this Supplementary Material (SM), we provide supporting derivations and technical details underlying the results of the main text. 
Appendix~A provides a derivation of the static structure factor $S(q)$ of the CF metal within the Halperin--Lee--Read (HLR) field theory~\cite{Halperin93} at the level of the random phase approximation (RPA), and shows that requiring a physically sensible response necessarily produces a $q^{3}\ln q$ correction to $S(q)$. Appendices B and C address finite-size aspects of our microscopic calculation of $S(q)$ using CF trial wave functions: (i) when mapping the angular-momentum-resolved structure factor $S_{l}$ obtained in spherical geometry to the planar $S(q)$, the mapping $q=\sqrt{l(l+1)}/R$ (with $R$ the sphere radius) yields optimal convergence; and (ii) the even-$l$ components of $S_{l}$ converge substantially faster to the thermodynamic limit than the odd-$l$ components. Appendix~D demonstrates the uniqueness of the Girvin--MacDonald--Platzman (GMP) density operator $\overline{\rho}_{\bm{q}}$~\cite{Girvin86} in terms of single CF particle--hole excitations in the $q \to 0$ limit, consistent with Kohn’s theorem~\cite{Kohn61}, and discusses its relation to parton and dipole constructions~\cite{Read98, Murthy02, Dong20}. Finally, Appendix~E shows that $S(q)$ for a noninteracting CF Fermi sea at filling $\nu$, yielding $S(q) \sim q^{2}/2 + (2k_{F}/3\pi)q^{3}$ as $q \to 0$, where $k_{F}=\sqrt{2\nu}$ is the Fermi wave vector and momentum $q$ is measured in units of $1/\ell_{B}$ with $\ell_{B}$ the magnetic length.

\section{Static Structure Factors from Field Theories}
\label{app: SSF_field_theories}
In this section, we review and discuss the predictions of the static structure factor $S(q)$ from the various field-theoretic perspectives of the CF metal that have been proposed over the years~\cite{Halperin93, Son15, wang_senthil_16}. In all such theories, the CFs are coupled to a dynamical gauge field that implements flux attachment. While the CFs themselves form a Fermi surface, their coupling to gapless gauge fluctuations gives rise to a non-Fermi-liquid state~\cite{Halperin93, Nayak94nfl, Polchinski94}. Despite considerable efforts over the years, a fully controlled theoretical description of the corresponding RG fixed point remains elusive. 

The electromagnetic response of the CF metal differs significantly from that of the Landau FL. First of all, in the simplest scenario, even if the CFs form a mean-field state that resembles a Fermi gas, they are not gauge-invariant. The physical response function $\Pi_{e}^{\mu\nu}(\omega,q)$ is determined by the Ioffe-Larkin rule~\cite{IoffeLarkin1989,LeeNagaosa1992,Halperin93}, $\Pi_{e}^{-1}=\Pi_{\textrm{CF}}^{-1}+\Pi_{\textrm{CS}}^{-1}$, where $\Pi_{\textrm{CF}}$ is the response function of the CFs, and $\Pi_{\textrm{CS}}$ is the Chern-Simons kernel. On the other hand, the coupling of the CF Fermi surface to a dynamical gauge field leads to Landau damping, while the associated gauge flux represents the physical electron density. 

A common approach to computing the electromagnetic response is the use of the random phase approximation (RPA). We begin by illustrating the idea using the Halperin-Lee-Read (HLR) theory for the half-filled LLL~\cite{Halperin93}. Here, because the flux of the emergent gauge field is tied to the electron density, the long-wavelength density response is related to the transverse gauge-field propagator at small-$q$. The standard RPA treatment yields
\begin{align}
\label{eq:RPA_SSF_Master_Formula}
S(\boldsymbol{q})=\tfrac{1}{n}\int\tfrac{\textrm{d}\omega}{2\pi}\tfrac{q^2/(2\pi p)^2}{\frac{q^2}{(2\pi p)^2}\Pi_{\textrm{CF}}^{\tau\tau}(\omega,q)^{-1}+\frac{q^2}{(2\pi p)^2}U(q)-\Pi_{\textrm{CF}}^{TT}(\omega,q)},
\end{align}
where $U>0$ corresponds to a repulsive interaction and the longitudinal and transverse components of the response function for a non-relativistic Fermi gas are given by
\begin{align}
  \Pi_{\textrm{CF}}^{\tau\tau}(\omega,q)&=\mathscr{D}_{F}\left(1-\frac{|\omega|}{\sqrt{\omega^{2}+(v_{F}q)^{2}}}\right),\nonumber\\
  \Pi_{\textrm{CF}}^{TT}(\omega,q)&=-\mathscr{D}_{F}\frac{|\omega|\sqrt{\omega^{2}+(v_{F}q)^{2}}-\omega^{2}}{q^{2}}. \label{eq:Pi_CF}
\end{align}
Here, $v_{F}$ is the Fermi velocity, $k_{F}$ denotes the Fermi momentum, and $\mathscr{D}_{F}=k_{F}/(2\pi v_{F})$ represents the density of states at the Fermi level. Following the convention in the main text, we have chosen to normalize $S(\bm{q})$ by the particle density $n = \pi k_F^2/(2\pi)^2$; in the $\nu=1/p$ state, setting $\ell_B=1$ fixes $k_F=\sqrt{2/p}$ since $n=1/2\pi \ell_B^2 p$. 

Now we may evaluate Eq.~\eqref{eq:RPA_SSF_Master_Formula}. If $\omega \gg q$ and $U(q)$ is less singular than $q^{-2}$, 

\begin{equation}
\label{eq:RPA_SSF_Master_Formula_large_w}
\begin{split}
    S(\boldsymbol{q})&\approx\tfrac{1}{n}\tfrac{q^2}{(2\pi p)^2}\int\tfrac{\textrm{d}\omega}{2\pi}\frac{1}{\tfrac{2}{(2\pi p)^2\mathscr{D}_F v_F^2}\omega^2+\tfrac{1}{2} \mathscr{D}_Fv_F^2 }\\
    &=\tfrac{q^2}{2}
\end{split}
\end{equation}
which is nothing but the contribution from the Kohn mode -- this is present despite the lack of manifest Galilean invariance in the HLR theory. To obtain the most relevant non-analytic correction, we take the opposite limit $\omega\ll q$. If we expand the denominator of the integrand in Eq. \eqref{eq:RPA_SSF_Master_Formula} to leading order in $\omega$, we have:
\begin{equation}
    \mathscr{D}_{F}v_F\frac{\abs{\omega}}{\abs{q}} \sim \frac{1}{(2\pi p)^2\mathscr{D}_F}q^2+\frac{1}{(2\pi p)^2}q^2U(q) 
\end{equation}
which yields a Landau damping pole after analytically continuing to real time. Unless $U(q)$ is long-range (i.e. singular as $q\to 0$), the inter-CF interaction will only renormalize the overall coefficient of the Landau damping rate, and we identify the scaling $\omega\sim iq^3$. If $U(q)\sim q^{-\eta}$, then we instead have $\omega\sim iq^{3-\eta}$. Evaluating the integral in this limit, 

\begin{align}
\label{eq:RPA_SSF_Master_Formula_Landau_damping}
&S_{L.D.}(\boldsymbol{q})=\nonumber\\
&\ \ \ \ \ \ \frac{q^3}{(2\pi p)^2 n}\int_0^{vq}\frac{\textrm{d}\omega}{\pi}\frac{1}{\frac{q^3}{(2\pi p)^2\mathscr{D}_F}+\frac{q^3}{(2\pi p)^2}U(q)+\mathscr{D}_{F}v_F\abs{\omega}}
\end{align}
where the upper limit of the integral is determined by some scale ($\propto v_{F}q$) where the $\omega\ll q$ approximation breaks down. First supposing $U(q)$ is regular at $q=0$, and using $\mathscr{D}_{F}v_F = (2\pi)^{-1}$, we have 

\begin{align}
\label{eq:RPA_SSF_Master_Formula_no_int}
S_{L.D.}(\boldsymbol{q})= \frac{1}{\pi\sqrt{p/2}} q^3\log(q_0/q)
\end{align}
Assuming $v=v_F$ for simplicity, $q_0 =  \frac{p}{\sqrt{1+\mathscr{D}_FU(0)}}$. 
If $U(q)= U_0q^{-\eta}$,
\begin{align}
\label{eq:RPA_SSF_Master_Formula_sing}
S_{L.D.}(\boldsymbol{q})= \frac{(2-\eta)}{\pi\sqrt{2p}} q^3\log(q_0/q)
\end{align}
where $q_0^{2-\eta} =(2\pi p)^2{\mathscr{D}_F v_F^2}/{U_0}$. Note that the dynamical exponent $z$ for the transverse gauge field fluctuations is $z=\min(3,3-\eta)$. Putting these contributions together, we have
\begin{align}
\label{eq:SSF_HLR_FINAL}
S(q)=\frac{1}{2}q^{2}+c_{U}q^{3}\log(q_0/q)+\mathscr{O}(q^{4}),
\end{align}
where the coefficient $c_{U}$ is determined by the inter-CF interaction $U(q)\sim q^{-\eta}$ according to
\begin{align}
c_{U}=\begin{cases}
1/\pi\sqrt{p/2}, & \eta\leq 0\\
(2-\eta)/\pi \sqrt{2p}, & 0<\eta<2\\
\end{cases}.
\end{align}
 
For the special case $U(q)=U_{0}q^{-2}$ ($\eta=2$), corresponding to a $\log(r)$ interaction in real space, the logarithmic non-analyticity in $S(q)$ is replaced by
\begin{equation}
S_{L.D.}(\bm{q})=\frac{1}{\pi\sqrt{2p}}\log\left(1+\frac{2\pi\sqrt{2p^3} v_F}{U_{0}}\right)q^{3}+\ldots,
\end{equation}
where the coefficient of the $q^{3}$ term depends on $v_{F}/U_{0}$. 
Although this appears to reproduce the microscopic $q^{3}$ behavior, it is unclear how this would reproduce the coefficient $2\sqrt{2/p}/3\pi$ obtained from our microscopic calculation. 

An additional problem arises in the $\omega \gg q$ limit; Eq.~\eqref{eq:RPA_SSF_Master_Formula} gives
\begin{equation}
S(\bm{q}) \approx \frac{1}{\sqrt{1+\frac{U_0}{2\pi p v_F}}}\frac{q^2}{2},
\end{equation}
which violates Kohn’s theorem as well as the bound on $q^2$ term\cite{Wu24}. This violation occurs within RPA for any interaction $U(q)\sim q^{-\eta}$ with $\eta\geq 2$, so RPA yields physically sensible results only for $0\leq\eta<2$, where the correction to $\overline{S}(q)$ is fixed as $\sim q^{3}\ln q$.

Lastly, in the Dirac CF theory proposed by Son \cite{Son15}, an analogous RPA calculation for emergent gauge field at $\nu=1/2$ gives:
\begin{align}
    S_{\rm Son}(\bm q) &= \frac{1}{n}\frac{q^2}{(4\pi)^2}\int \frac{d\omega}{2\pi} \frac{1}{\frac{q^2}{(4\pi)^2} U(q)-\Pi^{TT}_{\rm CF}}
\end{align}
where $\Pi^{TT}_{\rm CF}$ is the same as Eq. \eqref{eq:Pi_CF}. In the limit $\omega \ll v_F q$, the non-analytic contribution to $S_{\rm Son}(\bm q)$ from Landau damping is the same as in the HLR theory. The $q^2/2$ is not present in the minimal version of the theory.

\section{From Sphere to Plane: Mapping the Static Structure Factor $S(q)$}
\label{app: S_L_plane_sphere_IQHE}

In this section, we begin by analyzing the asymptotic behavior of the static structure factor $S_{l}$ for the $\nu=1$ integer quantum Hall state, in order to establish its thermodynamic limit and clarify its connection to the planar structure factor $S(q)$. We show that the mapping $q=\sqrt{l(l+1)}/R$, where $R$ is the radius of the Haldane sphere, provides the optimal mapping, with $S_{l}$ converging to the well-known result $S(q)=1-e^{-q^{2}/2}$ for all $q \in \mathbb{R}$ whereas the conventionally used mapping $q=l/R$ fails to do so as $q \to 0$. We also show with the mapping $q=\sqrt{l(l+1)}/R$, our microscopic results based on the CF trial wavefunctions for the CF metal states show much better convergence as a function of $N$ than the mapping $q=l/R$, again suggesting the former to be optimal.

\subsection{The $\nu=1$ IQH state}

The $\nu=1$ integer quantum Hall state of $N$ electrons on the sphere occurs at monopole strength $2Q=N-1$, since the LLL contains $2Q+1$ orbitals. For this state, the static structure factor $S_{l}$ is given by~\cite{Dora24}  
\begin{equation}
    S_{l} = 1-(2Q+1)
    \begin{pmatrix}
        Q & Q & l \\
        -Q & Q & 0 \\
    \end{pmatrix}^{2}.
\end{equation}

We are interested in how $S_{l}$ approaches its planar counterpart $S(q)=1-e^{-q^{2}/2}$~\cite{Girvin86} in the thermodynamic limit, i.e., as $N=(Q-1)/2\to\infty$. To proceed, note that the Wigner-$3j$ symbol above can be written in terms of $\Gamma$ functions as~\cite{Varshalovich88}
\begin{equation}
    \begin{pmatrix}
        Q & Q & l \\
        -Q & Q & 0 \\
    \end{pmatrix}^{2}
    = \frac{\Gamma(2Q+1)^2}{\Gamma(2Q-l+1)\,\Gamma(2Q+l+2)}.
\end{equation}

Substituting $Q=(N-1)/2$ gives  
\begin{equation}\label{eq:gamma-form}
    S_{l} = 1 - N \,\frac{\Gamma^{2}(N)}{\Gamma(l)\,\Gamma(N+l+1)}.
\end{equation}

Let us assume that there exists a unique optimal mapping between an angular momentum $l$ on the sphere of radius $R=\sqrt{Q}$ and a planar momentum $q$. On general grounds, for large $q$ we expect $l\sim qR$. Thus, the arguments of the $\Gamma$ functions in Eq.~\eqref{eq:gamma-form} scale linearly with $N$ (or $Q$) as $N\to\infty$, which allows us to use Stirling’s expansion,  
\begin{equation}
\begin{split}
    \Gamma(x) &\sim \exp\!\left[-x+x\log x -\tfrac12\log x \right.\\
    &\left.+\tfrac12\log(2\pi)+\tfrac{1}{12x}+O(1/x^3)\right].
\end{split}
\end{equation}

Applying this to Eq.~\eqref{eq:gamma-form} yields the asymptotic expansion  
\begin{align*}
    \lim_{N\to\infty} S_{l}
    &= 1 - \exp\!\left[-\frac{l(l+1)}{N} - \frac{l^{2}(l+1)^{2}}{6N^{3}} + O\!\left(\tfrac{1}{N^{4}}\right)\right].
\end{align*}

Hence, with the mapping
\[
q^{2}=\frac{2l(l+1)}{N}=\frac{l(l+1)}{Q}+O\!\left(\tfrac{1}{Q^{2}}\right)
\;\;\Leftrightarrow\;\;
q=\frac{\sqrt{l(l+1)}}{R},
\]
we find that $S_{l}$ converges to the planar structure factor in the thermodynamic limit for all values of $q$ (or $l$). By contrast, the conventional mapping $q=l/R$ exhibits significant deviations from the expected behavior as $q,l\to 0$.

\subsection{The CF metal states}  
\begin{figure}
    \includegraphics[width=\columnwidth]{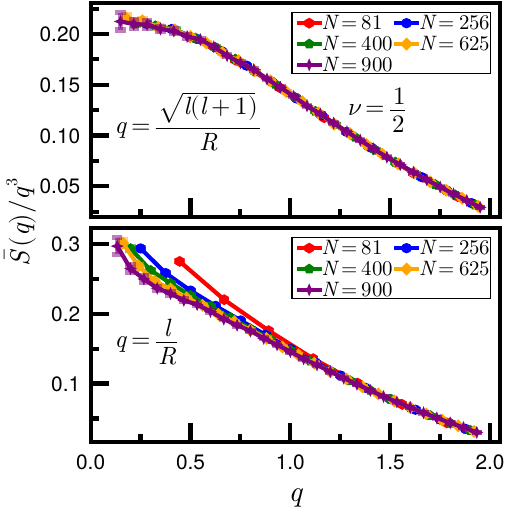}
    \caption{Comparison of two mappings between spherical and planar momenta for the CF metal state at $\nu=1/2$, shown for various system sizes $N$. \textit{Top:} results using $q=\sqrt{l(l+1)}/R$, which converge rapidly to the planar limit as $q\to 0$. \textit{Bottom:} results using the conventional mapping $q=l/R$, which display noticeable deviations even between $N=400$ and $N=900$. We see similar trends at $\nu=1,\,1/3,$ and $1/4$.}
    \label{fig:CFmetal_mapping}
\end{figure}
Figure~\ref{fig:CFmetal_mapping} shows our results for $\overline{S}(q)$ for the CF metal state at $\nu=1/2$, plotted for different system sizes $N$ using the two candidate mappings: $q=\sqrt{l(l+1)}/R$ (with $R=\sqrt{Q}=\sqrt{N-1}$) and $q=l/R$, restricted to even-$l$. As is evident, the mapping $q=\sqrt{l(l+1)}/R$ yields much better convergence to the planar limit as $q\to 0$, while the $q=l/R$ mapping exhibits sizable deviations even for system sizes as large as $N=900$. Similar behavior is observed at $\nu=1,\,1/3,$ and $1/4$, further supporting the conclusion that $q=\sqrt{l(l+1)}/R$ is the optimal mapping.

\section{Even- vs.~Odd-$l$ convergence of $\ov S_l$ for the CF metal states}
\label{app:even-odd}
\begin{figure}
    \centering
    \includegraphics[width=\linewidth]{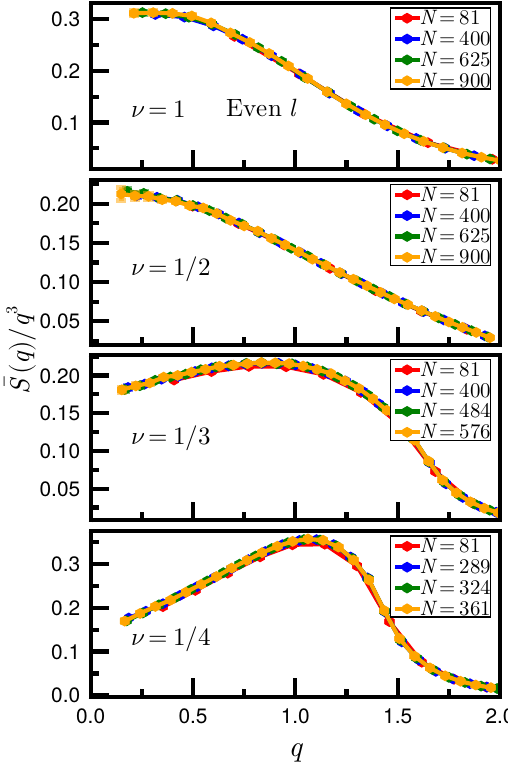}
    \caption{Projected static structure factor $\overline{S}(q)/q^{3}$ as a function of $q$ for CF metals at fillings $\nu=1,\,1/2,\,1/3,$ and $1/4$, shown for various system sizes $N$. The calculations are performed in spherical geometry, where we compute $S_{l}$, the correlation function of angular-momentum-resolved density operators, and then map to the planar structure factor via $q=\sqrt{l(l+1)}/R$ with $R$ the sphere radius. Here we show results for even-$l$. Converged behavior is evident already for $N\approx 81$, though the accessible momentum range is narrow. The wave-vector $q$ is measured in units of $1/\lB$, where $\lB$ is the magnetic length.}
    \label{fig:even-convergence}
\end{figure}

\begin{figure}
    \centering
    \includegraphics[width=\linewidth]{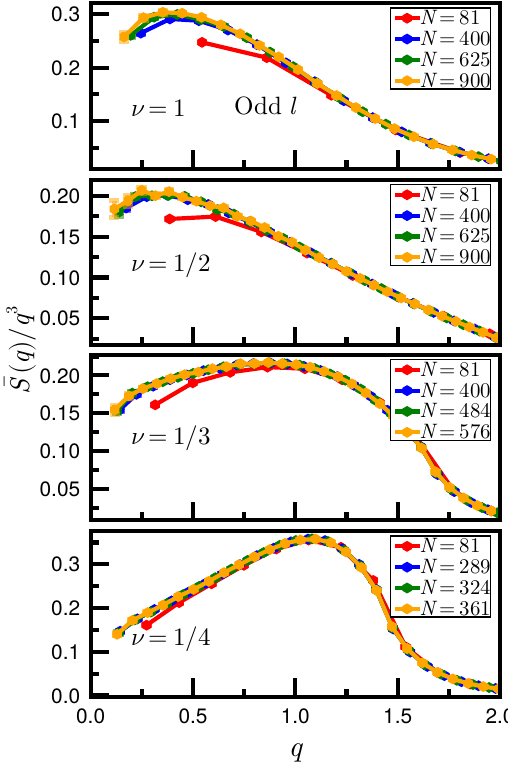}
    \caption{Same as Fig.~\ref{fig:even-convergence}, but for odd-$l$ values. In this case, strong finite-size effects persist, with noticeable deviations even for $N=900$.}
    \label{fig:odd-convergence}
\end{figure}

In this section, we present $\ov S_{l}$ for even and odd $l$ separately for the CF metals at $\nu=1,1/2,1/3,$ and $1/4$ in Fig.~\ref{fig:even-convergence} and Fig.~\ref{fig:odd-convergence}, respectively. As is evident, the even-$l$ values exhibit rapid convergence with system size $N$, while the odd-$l$ values converge much more slowly. For this reason, we restrict our analysis to the even-$l$ data, as they more faithfully capture the thermodynamic limit.

\section{Representations of the LLL charge density operator}
\label{app:GMP_MOTIVATION}

In this section, we demonstrate that in the $q \to 0$ limit, 
the GMP operator $\overline{\rho}_{\bm{q}}$ admits a unique 
representation in terms of single CF particle--hole excitations: 
$
\overline{\rho}_{\bm{q}} = \imath \sum_{\bm{k}} (\bm{q} \times \bm{k}) \,
c^{\dagger}_{\bm{k}+\bm{q}} c_{\bm{k}} .
$

In the main text we proposed that the projected density operator $\overline{\overline{\rho}}(\bm{q})$ can be represented as  
\begin{equation}\label{eq:rho-small-q-APP}
\oovv{\rho}(\bm{q})=\sum_{\bm{k}} f(\bm{q},\bm{k})\, c^\dagger_{\bm{k}+\bm{q}} c_{\bm{k}},
\end{equation}
where $c^\dagger_{\bm{k}}$ and $c_{\bm{k}}$ are CF creation and annihilation operators at zero magnetic field (i.e., in the CF metal phase) with the form factor $f(\bm{q},\bm{k})$ constrained to behave as  
\[
f(\bm{q},\bm{k}) \;\xrightarrow{q\to 0}\; i\ell_B^2\,(\bm{q}\times\bm{k}).
\]  
Assuming that the the CF operators obey canonical anti-commutation relations, we now show that this form of $f$ is uniquely mandated by the GMP algebra and Kohn’s theorem.  

First, we must have $f(\bm{0},\bm{k})=0$ in order to treat the ground state as a noninteracting Fermi sea of $c$-operators. If instead $f(\bm{0},\bm{k})=f_0(\bm{k})\neq 0$, the projected static structure factor would behave as $\overline{S}(\bm q\to0)\propto |f_0(k_F)|^2|\bm q|$, in violation of Kohn’s theorem.  

Next, the requirement that $\overline{\overline{\rho}}(\bm q)$ satisfies the GMP algebra implies that $f$ obeys the functional equation  
\begin{equation}
\begin{split}
    \label{eq:GMP_FUNCTIONAL_EQ}
    &\frac{f(\bm{q},\tfrac{\bm{q}'}{2}+\bm{k})\,f(\bm{q}',-\tfrac{\bm{q}}{2}+\bm{k})
    -f(\bm{q},-\tfrac{\bm{q}'}{2}+\bm{k})\,f(\bm{q}',\tfrac{\bm{q}}{2}+\bm{k})}
    {f(\bm{q}+\bm{q}',\bm{k})}
    \\&=2i \sin\!\Bigl(\tfrac{\bm{q}\times \bm{q}'}{2}\Bigr).
\end{split}
\end{equation}

To analyze the small-$q$ limit, let us denote  
\[
\partial_{q_\alpha} f(\bm q,\bm k)\big|_{q\to 0}=A_\alpha(\bm k), 
\quad 
\partial_{k_\alpha}A_\beta(\bm k)=B_{\alpha\beta}(\bm{k}).
\]
(We ignore the $q_{\alpha}q_{\beta} \partial_{q_{\alpha}q_{\beta}}f(\bm{q}, \bm{k})\vert_{q \to 0}$ like contributions, as they are not relevant at lowest order in $q, q^{\prime}$.)
Expanding Eq.~\eqref{eq:GMP_FUNCTIONAL_EQ} for $q,q'\to 0$ yields
\begin{equation}\label{eq:small-q-qprime}
(\bm{q}\cdot \bm{B}\cdot \bm{q}')(\bm{q}'\cdot \bm{A})-(\bm{q}'\cdot \bm{B}\cdot \bm{q})(\bm{q}\cdot \bm{A})
= i (\bm q\cdot \bm A + \bm q'\cdot \bm A)(\bm q \times \bm q').
\end{equation}
Setting $\bm q'=\lambda \bm q$ gives
\[
(\lambda^2-\lambda)(\bm q\cdot \bm B\cdot \bm q)(\bm q\cdot \bm A)=0,
\]
which implies $\bm q\cdot \bm B\cdot \bm q=0$. Hence $\bm B$ must be antisymmetric: $\bm q\cdot \bm B\cdot \bm q'= -\bm q'\cdot \bm B\cdot \bm q$. Substituting into Eq.~\eqref{eq:small-q-qprime}, we obtain
\[
\bm q\cdot \bm B\cdot \bm q' = i\, \bm q\times \bm q',
\qquad \implies \quad B_{\alpha\beta}= i\,\epsilon_{\alpha\beta}.
\]

It follows that  
\[
A_\alpha(\bm k)= i\epsilon_{\alpha\beta}k_\beta + c_\alpha,
\]
and thus, in general,  
\begin{equation}
f(\bm q,\bm k)\big|_{q\to 0} = c_\alpha q_\alpha + i\,\epsilon_{\alpha\beta}q_\alpha k_\beta.
\end{equation}
The first term, proportional to $c_\alpha$, breaks rotational invariance; hence in the isotropic CF metal we must set $c_\alpha=0$. Therefore we conclude that, uniquely,  
\begin{equation}
f(\bm q,\bm k)= i\lB^{2} \bm q\times \bm k \qquad (q\to 0).
\end{equation}

\subsection{Relation to parton constructions}
Now, following \cite{Murthy03, Murthy15, Dong20}, we would like to motivate the form of the operator in \ref{eq:GMP_charge_density_operator} which obeys the GMP algebra exactly. First, we recall that within the Pasquier-Haldane parton construction for bosons at $\nu=1$ \cite{Pasquier98, Read98}, we can define two operators

\begin{equation}
\begin{split}
\label{eq:PH_GMP}
\ov{\rho}_L(\bm{q}) &= e^{-q^2/4}\ov{ \overline{\rho}}_L(\bm{q})=e^{-q^2/4}\sum_k e^{i\bm{q}\times\bm{k}/2}c^\dag_{\bm{k}+\bm{q}}c_{\bm{k}}\\
\ov{\rho}_R(\bm{q}) &= e^{-q^2/4}\ov{\overline{\rho}}_R(\bm{q})=e^{-q^2/4}\sum_k e^{-i\bm{q}\times\bm{k}/2}c^\dag_{\bm{k}+\bm{q}}c_{\bm{k}}\\
\end{split}
\end{equation}
where $\ov{\rho}_L$ is the physical density (with $\ov{\overline{\rho}}_L$ obeying the GMP algebra) and $\ov{\rho}_R(\bm{q}\neq 0)$ annihilates physical states, with $\ov{\overline{\rho}}_R$ obeying the GMP algebra with the opposite sign. As $\ov{\rho}_R(\bm{q}\neq 0)$ annihilates physical states, we may add and subtract it at will from other operators acting on physical states. 

Accordingly, we can come up with two representations of the physical charge density that obey the GMP algebra. They are $\overline{\rho}_{e} = \overline{\rho}_L(\bm{q})$ and $\overline{\rho}_{d} = \overline{\rho}_L(\bm{q}) - \overline{\rho}_R(\bm{q})$, the latter of which is precisely Eq.~\eqref{eq:GMP_charge_density_operator}. Each of these representations implies a different assignment of quantum numbers to the $c$-fermions in the $q\rightarrow 0$ limit -- if we choose $\overline{\rho}_e$, then the $c$-fermions carry unit charge, and if we choose $\overline{\rho}_d$, then the $c$-fermions are interpreted as dipoles (since $\overline{\rho}_d(\bm{q}\rightarrow 0)$ coincides with Eq.~\eqref{eq:rho-small-q}). 

Then, the next step in any parton construction is to perform a standard Hartree-Fock mean-field theory in the $c$-operators, supposing that the Hamiltonian takes a generic form $\hat{H} = \sum_k U(\bm{q})e^{-q^2/2}\overline{\rho}(\bm{q})\overline{\rho}(-\bm{q})$. One can find two different mean-field ground states, depending on whether we choose the representation $\overline{\rho} = \overline{\rho}_e$ or $\overline{\rho} = \overline{\rho}_d$. Neglecting the possibility of pairing, the mean-field ground state will be a circular Fermi sea of the $c$-fermions with $k_F$ set by Luttinger's theorem and a dispersion set by the interaction $U(q)$. 

Neither mean-field ground state respects the gauge constraints enforced on this parton approach, but we may still ask whether these mean-field ground states capture any of the essential physics. If we elect to represent $\overline{\rho} = \overline{\rho}_e$, the corresponding mean-field ground state will have a projected static structure factor $\overline{S}(\bm{q}) = \frac{1}{N}\avg{\overline{\rho}_e(\bm{q})\overline{\rho}_e(-\bm{q})} \propto \abs{q}$, in violation of Kohn's theorem \cite{Kohn61}. However, as shown in Appendix \ref{app: dipole_Fermi_sea}, choosing $\overline{\rho} = \overline{\rho}_d$ yields Eq.~\eqref{eq:dipolar_planar_SSF_final} as the mean-field result for $\ov S(\bm{q})$, which is consistent with Kohn's theorem. According to this discussion, Eq.~\eqref{eq:dipolar_planar_SSF_final} can be thought of as a result from parton mean-field theory, at least for the $\nu=1$ bosons.

\section{Fermi sea of non-interacting dipolar CFs}
\label{app: dipole_Fermi_sea}

Using the expression \ref{eq:rho-small-q} for the charge density, we assume that the ground state is a Fermi gas of the $c$-fermions and compute the static structure factor $\ov{S}(\bf{q})$ at small momentum. We assume no renormalization from the Fermi liquid theory, nor do we consider the effects of any gauge fluctuations. 

With these assumptions, the projected static structure factor $\ov{S}(q)$ in the planar geometry can be computed directly. The calculation is a straightforward adaptation of a textbook derivation~\cite{Giamarchi04}. We obtain (for $q >0$):
\begin{equation}
\begin{split}
\label{eq:dipolar_planar_SSF_step_1}
\ov{S}(q)&=\frac{1}{n}\bigl\langle \overline{\rho}(\bm{q})\overline{\rho}(-\bm{q}) \bigr\rangle\\
&= \frac{1}{2n}\int \frac{d^2k}{(2\pi)^2}(\bm{q}\times\bm{k})^2 \\
&\quad\times \Bigl[\vartheta(-\xi_{\bm{k}-\tfrac{1}{2}\bm{q}})-\vartheta(-\xi_{\bm{k}+\tfrac{1}{2}\bm{q}})\Bigr] \\
&\quad\times \mathrm{sgn}\left(\xi_{\bm{k}-\tfrac{1}{2}\bm{q}}-\xi_{\bm{k}+\tfrac{1}{2}\bm{q}}\right),
\end{split}
\end{equation}
where $\vartheta(x)$ is the Heaviside step function and $\xi_{\bm{k}}$ the CF dispersion. Since this expression depends only on the geometry of the filled Fermi sea, the specific form of $\xi_{\bm{k}}$ is unimportant; only rotational symmetry is assumed.
In the limit $\bm{q}\to 0$, expanding to leading order yields
\begin{align*}
\bigl\langle \overline{\rho}(\bm{q})\overline{\rho}(-\bm{q}) \bigr\rangle
&=\frac{1}{2}\int \frac{d^2k}{(2\pi)^2}(\bm{q}\times\bm{k})^2 \\
&\quad\times \Bigl[\bm{q}\cdot\bm{v}_{\bm{k}}\delta(-\xi_{k}) + O(q^3)\Bigr] \\
&\quad\times \mathrm{sgn}\bigl(-\bm{q}\cdot \bm{v}_{\bm{k}}\bigr),
\end{align*}
with $\bm{v}_{\bm{k}}=\nabla \xi_{k}$.
Evaluating the integral on the Fermi surface gives
\begin{align*}
\bigl\langle \overline{\rho}(\bm{q})\overline{\rho}(-\bm{q}) \bigr\rangle
&= \frac{1}{2}\int \frac{kdkd\theta}{(2\pi)^2} q^3 k^2 \sin^2\theta \cos\theta\delta(k-k_F) \\
&\quad\times \mathrm{sgn}(-qv_F\cos\theta) \\
&= \frac{1}{6\pi^2} k_F^3 q^3.
\end{align*}
Normalizing by the CF density $n_{\rm CF} =\pi k_F^2/(2\pi)^2$, which follows from the Luttinger relation, we arrive at
\begin{equation}
\begin{split}
\label{eq:dipolar_planar_SSF_final}
\ov{S}_{\rm dipole}(q)
&= \frac{1}{n}\bigl\langle \rho(\bm{q})\rho(-\bm{q}) \bigr\rangle \\
&= \frac{2k_F}{3\pi}q^3 + O(q^4) \\
&= \frac{2\sqrt{2}}{3\pi}\frac{1}{\sqrt{p}}q^3 + O(q^4),
\end{split}
\end{equation}
where we used $k_F=\sqrt{4\pi n_{\mathrm{CF}}}=\sqrt{2\nu}$, with $\nu=1/p$. Thus, the long-wavelength expansion yields the result quoted in the main text, demonstrating that the dipolar picture  reproduces the microscopic $\overline{S}(\bm{q})$ at small momentum.
\end{document}